\documentclass[fleqn,10pt]{SelfArx} 

\usepackage{lipsum} 
\usepackage{listings}
\usepackage[utf8]{inputenc}

\usepackage{color}

 \usepackage{afterpage}

\definecolor{codegreen}{rgb}{0,0.6,0}
\definecolor{codegray}{rgb}{0.5,0.5,0.5}
\definecolor{codepurple}{rgb}{0.58,0,0.82}
\definecolor{backcolour}{rgb}{0.95,0.95,0.92}

\lstdefinestyle{mystyle}{
    backgroundcolor=\color{backcolour},   
    commentstyle=\color{codegreen},
    keywordstyle=\color{magenta},
    numberstyle=\tiny\color{codegray},
    stringstyle=\color{codepurple},
    basicstyle=\footnotesize,
    breakatwhitespace=false,         
    breaklines=true,                 
    captionpos=b,                    
    keepspaces=true,     
    numbers=left,		
    numbersep=5pt,                  
    showspaces=false,                
    showstringspaces=false,
    showtabs=false,                  
    tabsize=2
}
 
\definecolor{color1}{RGB}{0,0,90} 
\definecolor{color2}{RGB}{0,20,20} 

\usepackage{hyperref} 
\hypersetup{hidelinks,colorlinks,breaklinks=true,urlcolor=color2,citecolor=color1,linkcolor=color1,bookmarksopen=false,pdftitle={Title},pdfauthor={Author}}

\JournalInfo{Accepted for publication in PASP} 
\Archive{} 

\PaperTitle{Hunting electromagnetic counterparts of gravitational-wave events using the Zwicky Transient Facility} 

\Authors{Shaon Ghosh\textsuperscript{1}*, Deep Chatterjee\textsuperscript{1},  David L. Kaplan\textsuperscript{1},  Patrick R. Brady\textsuperscript{1},  Angela Van Sistine\textsuperscript{1}} 
\affiliation{\textsuperscript{1}\textit{Department of Physics, University of Wisconsin Milwaukee}} 
\affiliation{*\textbf{Corresponding author}: ghosh4@uwm.edu} 

\newcommand{\keywordname}{Keywords} 

\Abstract{ 
  Detections of coalescing binary black holes by LIGO have opened a new window of transient astronomy.
  With increasing sensitivity of LIGO and
  participation of the Virgo detector in Cascina, Italy, we
  expect to soon detect coalescence of compact binary systems with
  one or more neutron stars. These are the prime targets for electromagnetic
  follow-up of gravitational wave triggers, which holds enormous promise of rich science. 
  However, hunting for electromagnetic counterparts of
  gravitational wave events is a non-trivial task due to the sheer
  size of the error regions, which could span hundreds of square degrees. 
 This may require deep observation with
  large field-of-view telescopes and/or use of galaxy catalogs. The Zwicky Transient facility (ZTF),
  scheduled to begin operation in 2017, is designed to cover
  such large sky-localization areas. In this work, we present the
  strategies of efficiently tiling the sky to facilitate the
  observation of the gravitational wave error regions using ZTF. To do
  this we used simulations consisting of 475 binary neutron
  star coalescences detected using a mix of two- and three-detector
  networks. Our studies reveal that, using two overlapping sets of ZTF
  tiles and a (modified) ranked-tiling algorithm, we can cover the gravitational-wave
  sky-localization regions with half as many pointings as a simple contour-covering 
  algorithm. We then incorporated the ranked-tiling strategy to study our ability to 
  observe the counterparts. This requires optimization of observation 
  depth and localization area coverage. Our results show that observation 
  in r-band with $\sim 600$ seconds of integration time per pointing 
  seems to be optimum for typical assumed brightnesses of electromagnetic counterparts, if we plan to 
  spend equal amount of time per pointing. However, our results also reveal 
  that we can gain by as much as $50\%$ in detection efficiency if we linearly
  scale our integration time per pointing based on the tile probability.}

\begin{document}

\flushbottom 

\maketitle 

\tableofcontents 

\thispagestyle{empty} 


\section*{Introduction} 

\addcontentsline{toc}{section}{Introduction} 

Sky-localization of gravitational wave (GW) events detected by LIGO-Virgo interferometers often cover hundreds of square degrees. The first direct detection of GW  from a coalescing binary black hole (BBH), GW150914, was localized in the sky ($90\%$ confidence interval) over an area of 630 square degrees \cite{Abbott:2016gcq} at the time of the alert. Events with a lower signal to noise such as GW151226, have larger sky-localization area, in excess of $800$ square degrees \cite{PhysRevX.6.041015}. The sky-localization for triggers with lower significance (which could be the case for non-BBH triggers)  could potentially be larger, often spanning over a thousand square degrees. The Zwicky Transient Facility (ZTF) \cite{Bellm:2014pia} with its $\sim 50$ square degrees field-of-view (FOV) would need tens of pointings to get a single observation over the $90\%$ credible region. We expect ZTF to reach a limiting magnitude of $r=20.5$ in 30 seconds. Kilonovae, which are EM emissions from $r$-process nucleosynthesis triggered by coalescing binaries of neutron star(s) \cite{1538-4357-507-1-L59}, continue to have light-curve models with a lot of uncertainties. Estimates of the absolute magnitudes of kilonovae range from $-12$ to $-15$ \cite{2041-8205-736-1-L21, 0004-637X-775-2-113}. For sources at 200 Mpc this corresponds to apparent magnitudes of $\sim 21.5$ to $24.5$ in the R-band. Thus If the desired depth is higher than the standard depth of $r=20.5$ for gravitational-wave events, much larger integration time could be required. Furthermore, a single observation may not be enough to identify transients. Observations at two different epochs, preferably with two different filters will most likely be necessary to get sufficient photometric information necessary to discern the actual electromagnetic (EM) counterpart of the gravitational wave triggers from a myriad of false positive \cite{0004-637X-814-1-25}. The first step towards such observation is an efficient sky-tiling method that will allow us to set up an observing strategy for any given event. In this paper, we present a sky-tiling method that is applicable for telescopes with wide fields of view and fixed tile coordinates, where, the fields at which the telescopes can point to are predefined. This is helpful for image subtraction required for transient identification. We used specifications of the ZTF telescope obtained from \cite{Bellm:2014pia} for this work. Note that this is a sky-tiling strategy study and therefore we did not take into account observing conditions such as geographic location of the telescope, the visibility conditions like phases of the moon and weather. Some of these aspects have been studied in the past \cite{Rana:2016crg, Srivastava:2016fyr} and others we are currently investigating. 

The structure of the paper is as follows: In Sec.~\ref{sec:tiling_strategies} we introduce the tiling strategy we propose where we compare its performance against a naive contour-covering strategy (Sec.~\ref{sec:cc_tiles}). We explain the various algorithms of implementing the tiling strategy (ranked-tiling strategy) in Sec.~\ref{sec:ranked_tiles}. In Sec.~\ref{sec:comparative_study} we present the implementation of these various strategies to analyze a large sample of sky-localization of gravitational wave simulations from \cite{Singer:2014qca}. Finally, having studied the various tiling strategies we use the optimal strategy to investigate the observing strategy where we optimize the depth of observation. This work is presented in Sec.~\ref{Sec:DepthCoverage}


\section{Tiling Strategies}
\label{sec:tiling_strategies}
Sky-localization information of the alerts from LIGO-Virgo comes in the form of HEALPix (Hierarchical Equal Area isoLatitude Pixelisation) sky-map \cite{0067-0049-226-1-10}. Each pixel has the probability of the gravitational-wave source being located at the center of the pixel, computed from the GW strain data of the interferometer. The observation of the sky-localization regions involves strategically tiling the sky, where each tile corresponds to the footprint of the field of view of the telescope. The localization probability contained by a single tile is given by the sum of all the pixels that lie within its boundaries. We describe two strategies for sky-tiling:

\subsection{The contour-covering tiles}
\label{sec:cc_tiles}
This is the most commonly used strategy when it comes to EM follow-up of GW triggers \cite{Kasliwal:2016uhu, 2041-8205-823-2-L33, doi:10.1093/mnras/stw1893}. It is also a sub-optimal one, especially when the tile coordinates are predefined on a fixed grid in the sky \cite{Ghosh:2015sxp}. However, given  the simplicity of the concept, we discuss this first. From the sky-maps one can construct the smallest $90\%$ confidence  interval for the source localization. If we enclose this region(s) \footnote{Note that the smallest $90\%$ confidence region might be multi-modal over the sky.} by contour(s), we can define a $90\%$ credible region resembling patch(es) in the sky. Contour-covering tiles (CC-tiles), as the name suggests, are the smallest number of tiles that are required to enclose this $90\%$ credible region. For ZTF, the tile coordinates are predefined in the sky, thus, the set of tiles that cover a given $90\%$ confidence interval is unique. Note that the CC-tiles will always cover more than $90\%$ localization probability due to finite size of the telescope's FOV. However, any additional localization that the CC-tiles cover is incidental and is most often not the best use of additional observation time or the ideal way to cover an additional area.

\subsection{The ranked tiles}
\label{sec:ranked_tiles}
\begin{figure}[ht]\centering
\includegraphics[width=\linewidth]{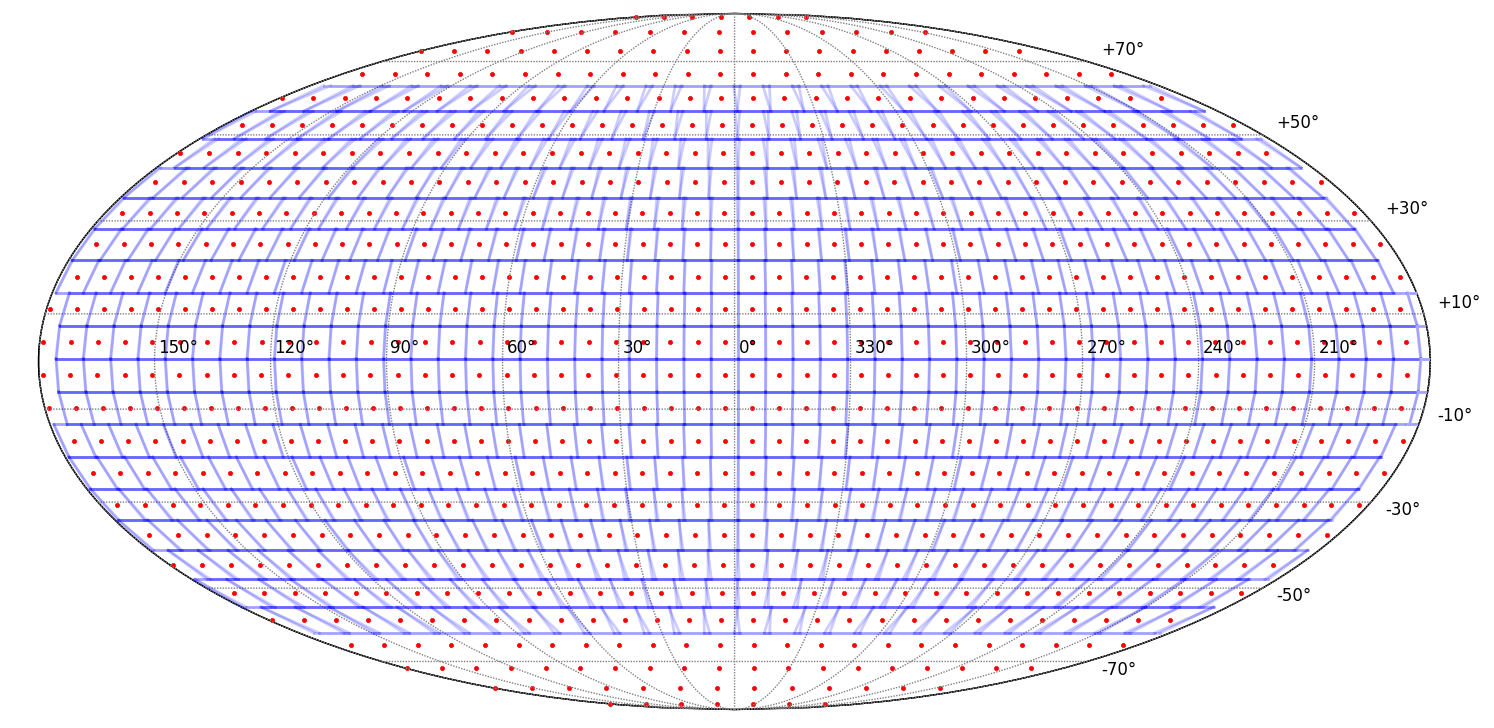}
\caption{Planned tile locations for ZTF: The red points are the tile centers and the blue squares are the tile boundaries. We do not show the tile boundaries for the tiles at large declination values (near the poles) as the boundaries (approximated as sides of trapezoids) get severely distorted near the poles.}
\label{fig:ZTF_Tiles}
\end{figure}

There exists a more natural way of tiling the sky that suits any particular telescope whose observing fields are predetermined. Instead of covering the contour, we can compute the sky-localization probability enclosed by all the tiles in the sky shown in Fig.~\ref{fig:ZTF_Tiles}. We then rank the tiles based on their localization probabilities and then choose from the top of these ranked tiles the set of tiles whose cumulative probability sum is $\geq 90\%$.
\\We implemented two distinct algorithms to do this:
\subsubsection{Tagging pixels to tiles}
Two observed facts about the ZTF tiles motivated this method. First, the ZTF tiles are placed in such a way (Fig.~\ref{fig:ZTF_Tiles}) that there are no gaps in the sky. This means that there are no orphan pixels, i.e., every HEALPix pixel will be within a tile. Second, ZTF's rectangular FOV on a spherical sky-surface means that each tile will always have some overlap with neighboring tiles, with the overlap getting larger near the celestial poles. Both of these features of the ZTF tiles help us to adopt an algorithm, where we loop through the pixels, and for each of them we search for the nearest tile center. We assign each pixel to its nearest tile center and then compute the accumulated probability for each tile. This leads to the required ranked tiles, which are the list of tiles in descending order of probability values contained within the tiles. Looping over the pixels ensures that we are not double counting pixels in the overlap regions.
\subsubsection{Greedy algorithm}
This is a modification of the algorithm above:
\begin{enumerate}
\item{For each tile, integrate the probability over the tile and sort the tiles based on these values.}
\item{Beginning with the highest value, include that tile in the list of those to cover and increment the summed probability}
\item{Zero out the probability for that whole tile so that the sky area is not repeated}
\item{Recompute the integrated probabilities with the updated sky map, resort and repeat step 2 until the required probability is reached.}
\end{enumerate}

\begin{figure}[ht]\centering
\includegraphics[width=\linewidth]{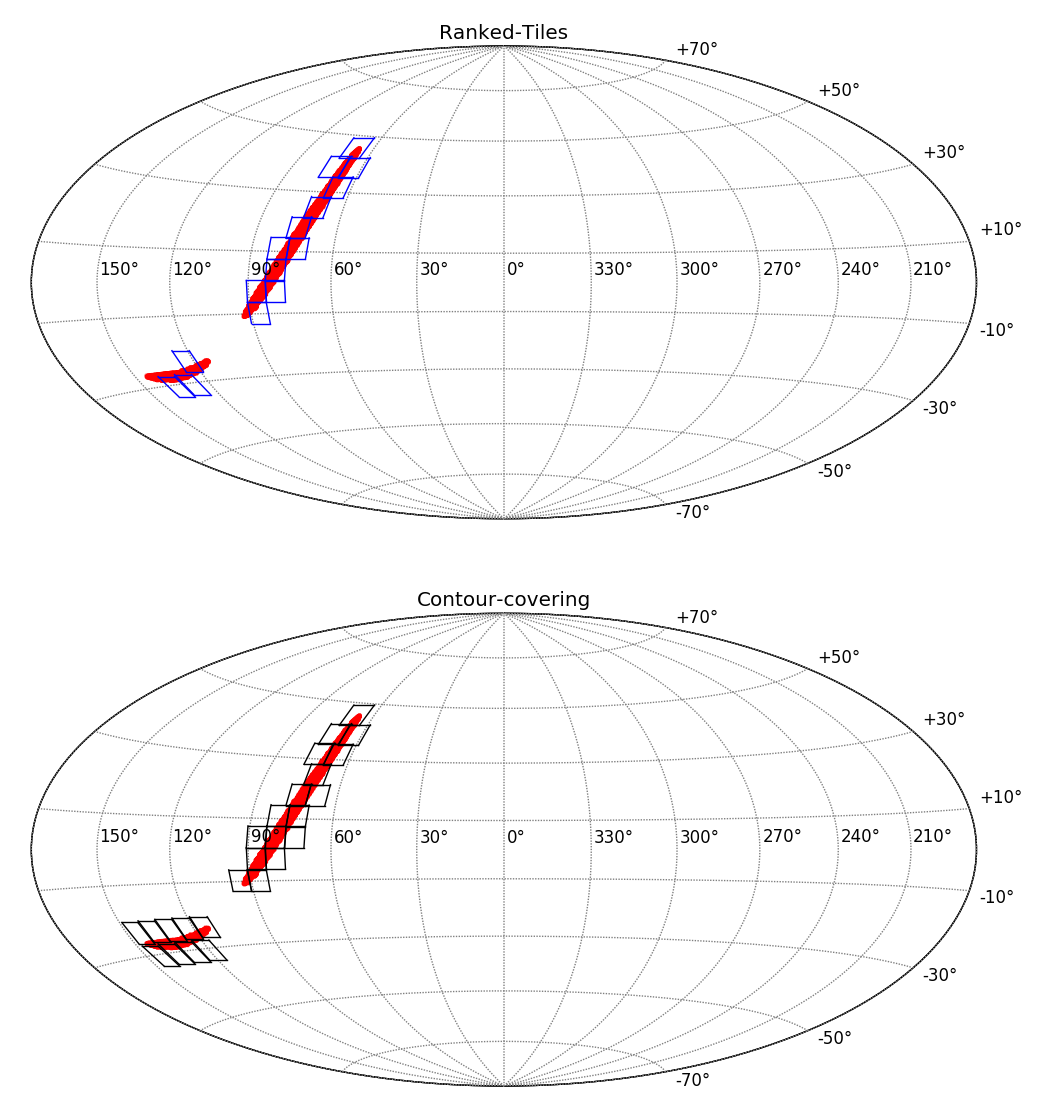}
\caption{Comparison between ranked tiles (top) and contour-covering tiles(bottom) for a simulated gravitational-wave sky-localization (event number 632720 from \cite{Singer:2014qca}). In this case we note that the 15 ranked tiles (blue) required to cover the $90\%$ confidence interval in sky-localization are a subset of the 26 contour-covering tiles (black).}
\label{fig:compare_tiling}
\end{figure}

Ranked tiling is however computationally slower than the contour-covering method in selecting the tiles required to cover $90\%$ localization. This is because in the contour covering method we know the location of the pixels of interest as shown in Sec.~\ref{sec:tiling_strategies} based on the contour of the confidence interval. If we can label these pixels (using a pixel ID for example), then all that is required beyond that is to find the tiles that have these labeled pixels. However, speed of the ranked tiling can be vastly improved by pre-computing which pixels belong to each tile. This mapping of the pixels to tiles on the predefined grid needs to be done once for a given telescope. For example, the densest (hence most computationally expensive) sky-maps require close to an hour for the computation of the ranked tiles. However, we achieve a speed up in the computation by $100$ times upon pre-computation, making the computation time requirement comparable to the contour-covering method. Thus, the present analysis uses the pre-computation mentioned above. Moreover, in the release of the code-base linked in Sec.~\ref{Sec:discussion}, we provide the user with the infrastructure required to conduct the pre-computation.
We show in Fig.~\ref{fig:compare_tiling} the performance of the CC-tiling and ranked-tiling strategies in generating the ZTF pointings for a simulated event.

\section{Comparing contour-cover and ranked tiles}
\label{sec:comparative_study}
\begin{figure}[ht]\centering
\includegraphics[width=\linewidth]{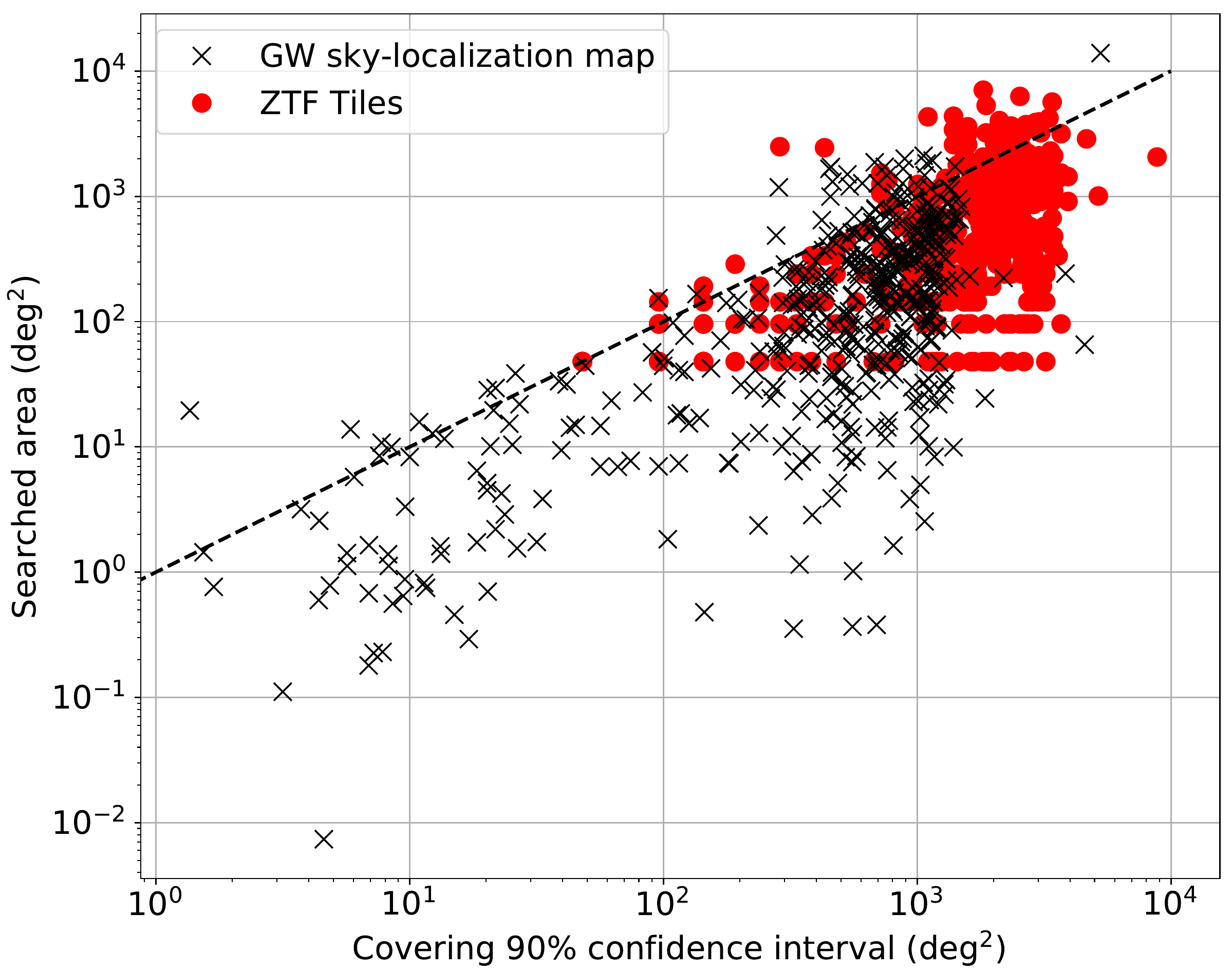}
\caption{Comparison of sky area searched to cover 90\% of the probability using a ranked-tiling strategy (horizontal axis) with the area searched to actually cover the simulated gravitational wave event (vertical axis).  This is shown for both the original location. The blue dots correspond to the original sky-localizations data obtained from the HEALPix sky-maps (blue dots) and the ZTF tiles (red stars).}
\label{fig:first2year_all_events}
\end{figure}

Having described the two methods of tiling (ranked and contour covering), we present the result of a study that was conducted over 475 simulated GW events. These events were obtained from the sky-localization study of binary neutron star coalescences from the simulation of 2016 LIGO-Virgo noise data \cite{Singer:2014qca}, which is a mix of two- and three-detector networks appropriate for early ZTF operations. In Fig.~\ref{fig:first2year_all_events}, we compare the area of the $90\%$ credible region covered by the ranked-tiling strategy with that needed to actually cover the simulated GW event location. If the sky-localizations are consistent, then $\sim 90\%$ of the events, depicted by red stars, should be below the black line. For comparison we also include the sky-localization searched area vs $90\%$ coverage in blue dots. This reflects the consistency of the sky-maps themselves. We find that $89.6\%$ of the black crosses, and $88.9\%$ of the red circles fall below the dashed black line. Thus, the localization probability enclosed in the GW sky-maps is consistent with the actual location of the source, and the ZTF tiles gives reasonably similar results (slightly worse due to discrete size FOV of the telescope which is significantly larger than the unit HEALPix pixel size).

\begin{figure}[ht]\centering
\includegraphics[width=\linewidth]{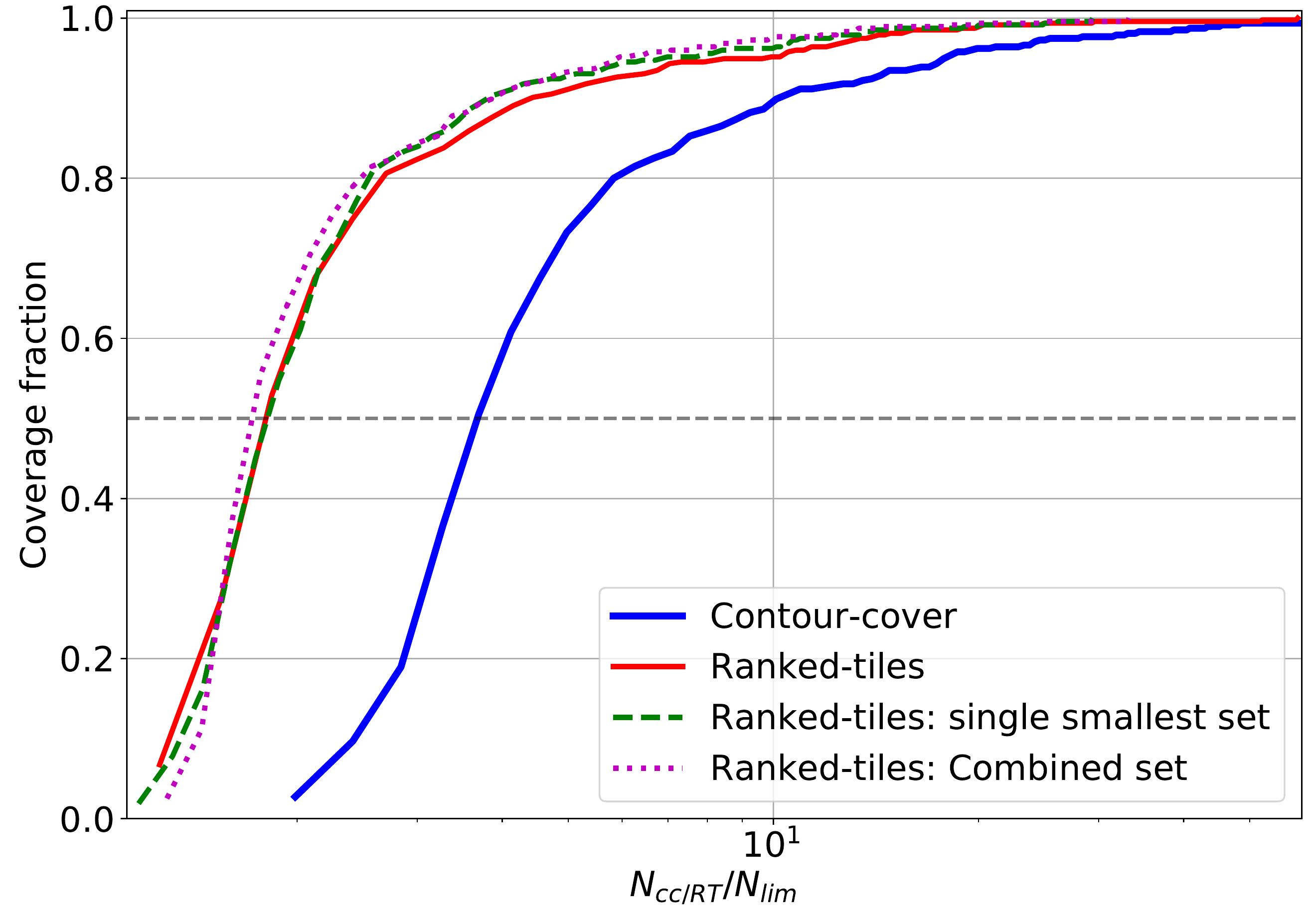}
\caption{Cumulative histogram of the ratio of $N_{X}$ to $N_{\rm lim}$ for different strategies and tile sets: contour-covering (solid blue), ranked tiling with one tile set (solid red), ranked tiling with the smallest of two sets (red dashed) and ranked tiling with a combined set (red dot-dashed). Ranked tiles with single smallest set shows the case where ranked-tiling algorithm was implemented for two sets of tiles and the smallest was chosen. Ranked tiles combined set is the case where we implemented the greedy ranked-tiling algorithm to obtain the ranked-tile set from the two sets of tiles.}
\label{fig:compare_tiling_allEvents}
\end{figure}

We then compared the ranked-tiling and contour-covering algorithms for all the 475 events using expected ZTF tiles. To quantify the comparison, we compute the number of tiles required by the two methods to cover the $90\%$ localization confidence interval for the ranked tiles ($N_{RT}$) and the $90\%$ credible region for the contour-covering tiles ($N_{CC}$). For each event we also define a packing-fraction, which is the limiting number of tiles required to cover the $90\%$ confidence interval $N_{\rm lim}$, 
\begin{equation}
\label{eq:packingFraction}
N_{\rm lim} = \frac{N_{\rm pix} \times A_{\rm pix}}{A_{\rm fov}}\,,
\end{equation}
where $N_{\rm pix}$ is the number of HEALPix pixels in the $90\%$ localization region, $A_{\rm pix}$ is area of each pixel, together giving the area of the $90\%$ localization region, and $A_{\rm fov}$ is the area of the
field-of-view of the telescope. Dividing $N_{CC}$ and $N_{RT}$ by $N_{\rm lim}$ provide a metric for comparison of the tiling methods with the minimal limiting case.  This quantity is plotted on the horizontal axis of Fig.~\ref{fig:compare_tiling_allEvents} where we histogram the result of the tiling by the two methods. Larger values of $N_{\rm X}/N_{\rm lim}$, where $X=CC$ or $RT$, indicate greater deviation from the limiting value and hence poorer coverage by the tiles. The median value of $N_{\rm X}/N_{\rm lim}$ for the ranked tiles is $1.91$, compared to $3.89$ for contour-covering tiles. This indicates that for the binary neutron star sky-localizations obtained from \cite{Singer:2014qca} the ranked-tiling strategy's performance is $\sim 2$ times better than the CC-tiling. Note that in \cite{Kasliwal:2013yqa} the factor $N_{CC}/N_{lim}$ was computed to be around $2.6$, although those authors used a different set of simulations and optimized set of CC-tiles. The CC-tiles in our study does not allow for any optimization and are closer to the strategy that will be adopted by ZTF. 

\begin{figure}[ht]\centering
\includegraphics[width=\linewidth]{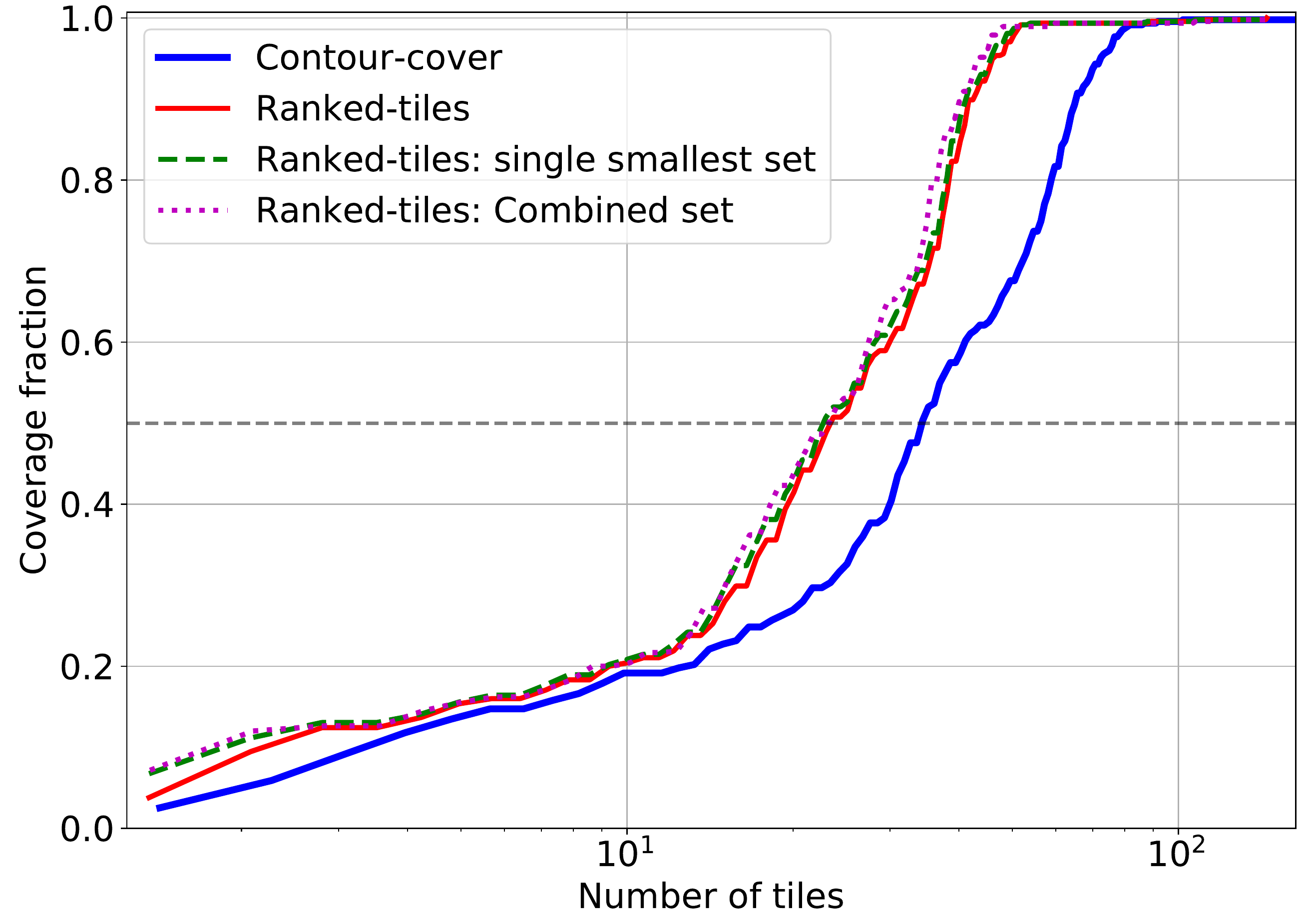}
\caption{Cumulative histogram of coverage fraction as a function of number of tiles covered according to various tiling methods. We note that all the ranked-tiling strategies give consistently better result than the contour-covering technique.}
\label{fig:compare_tiling_allEventsNumTiles}
\end{figure}

ZTF is expected to include two sets of tiles, with the second set offset from the first by half a tile in Right Ascension. We conducted the same analysis as above for the two tile-sets. For each event we conducted the study using the two sets independently and chose the one that gives the smaller number of tiles, with a median $N_{RT}/N_{\rm lim}$ of 1.86: just a $2.6\%$ improvement over the basic ranked-tiles case. If we apply the greedy algorithm to select tiles from both the sets simultaneously,  the value of $N_{RT}/N_{\rm lim}$ becomes 1.79: a $6.7\%$ improvement over the basic ranked-tiles case.  It is clear that there is limited room for further improvement, as the ratio is already getting close to 1. We also show the cumulative histogram of the number of tiles required to reach a given coverage fraction for all the method in Fig. \ref{fig:compare_tiling_allEventsNumTiles}. We note that the simple ranked-tiling method gives reasonably similar results using a single set to those incorporating the two tile sets to create the ranked tiles. 

\begin{table}[hbt]
\caption{Summary of the results of the tiling of gravitational wave events.}
\label{tab:summary}
\centering
\begin{tabular}{llr}
\toprule
\toprule
\cmidrule(r){1-2}
Tiling strategy & Median $N_{X}/N_{\rm lim}$\\
\midrule
Contour cover & 3.89 \\
Contour cover (optimized)& 2.60 \\
Ranked-tiling & 1.91 \\
Ranked-tiling (smallest )& 1.86 \\
Ranked-tiling (combined)& 1.79 \\
\bottomrule
\end{tabular}
\label{tab:tiling_startegy_performance}
\end{table}	
In Table~\ref{tab:summary} we summarize the result of this study. The first column shows the tiling strategy, where ranked-tiling (smallest) is the strategy where we use two sets of tiles and find the ranked tiles for both the sets independently, then we choose for each event the set with the minimum number of tiles. Ranked-tiling (combined) is the strategy where the two sets of tiles were used together to find a single set of non-overlapping ranked tiles for each event. Contour covering (optimized) is the number obtained from \cite{Kasliwal:2013yqa}. The second column shows the median of $N_{X}/N_{\rm lim}$ for each method. Note that this is a separate simulation of binary neutron star coalescence, conducted with LIGO-Virgo design sensitivity noise curve. Thus, this number is purely for reference of previous studies and should not be compared quantitatively with the other numbers in this study.

\section{ZTF depth vs coverage}
\label{Sec:DepthCoverage}
\begin{figure}[ht]\centering
\includegraphics[width=\linewidth]{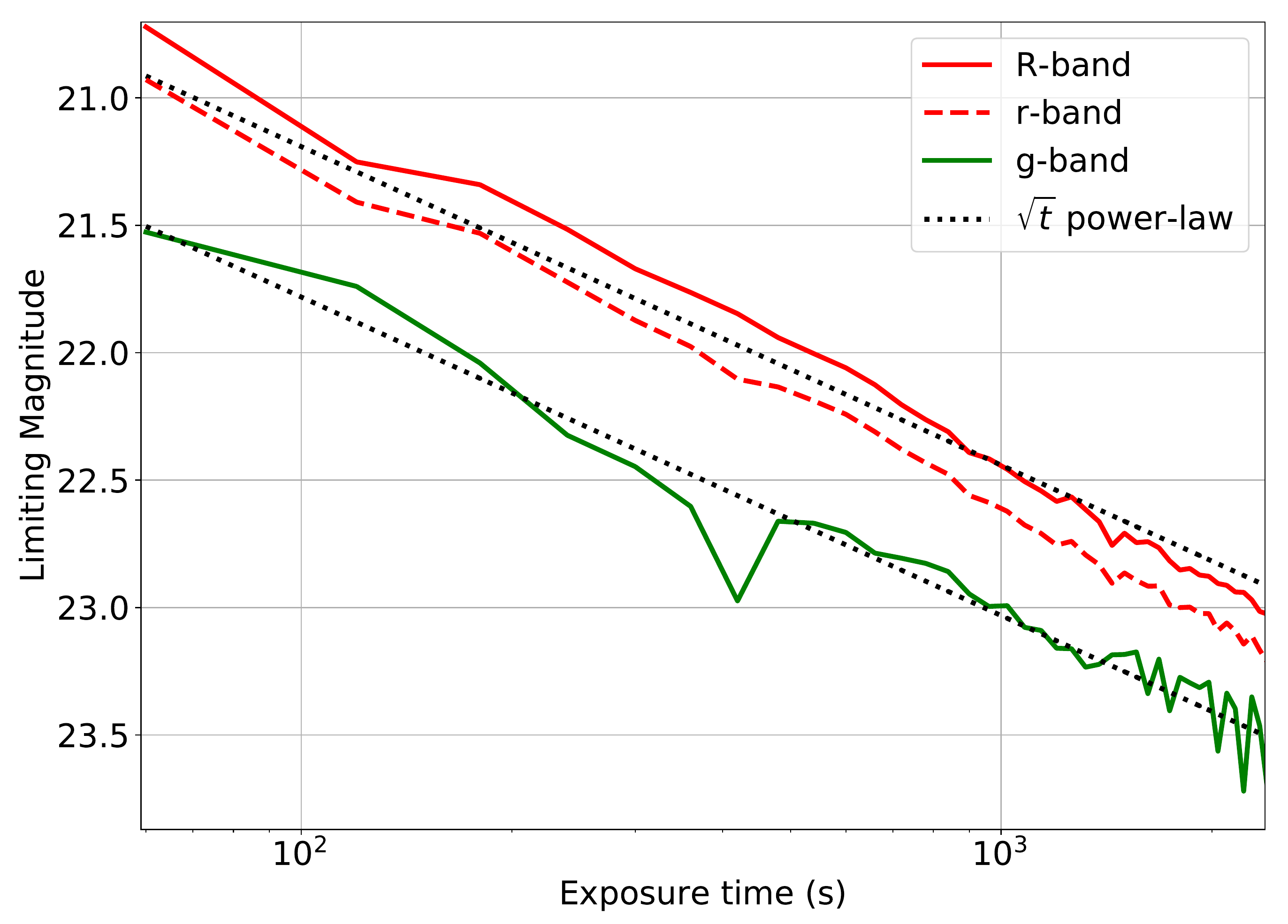}
\caption{Assumed integration time vs.~limiting magnitude for ZTF, based on the g- and R-band PTF data. PTF g-band and R-band are shown by the green and red solid lines respectively. We converted the R-band limiting magnitude to r-band limiting magnitude using Lupton (2005).  We also show the expected $S/N\sim\sqrt{t}$ behavior for comparison.}
\label{fig:time_limiting_mag}
\end{figure}

\begin{figure}[ht]\centering
\includegraphics[width=\linewidth]{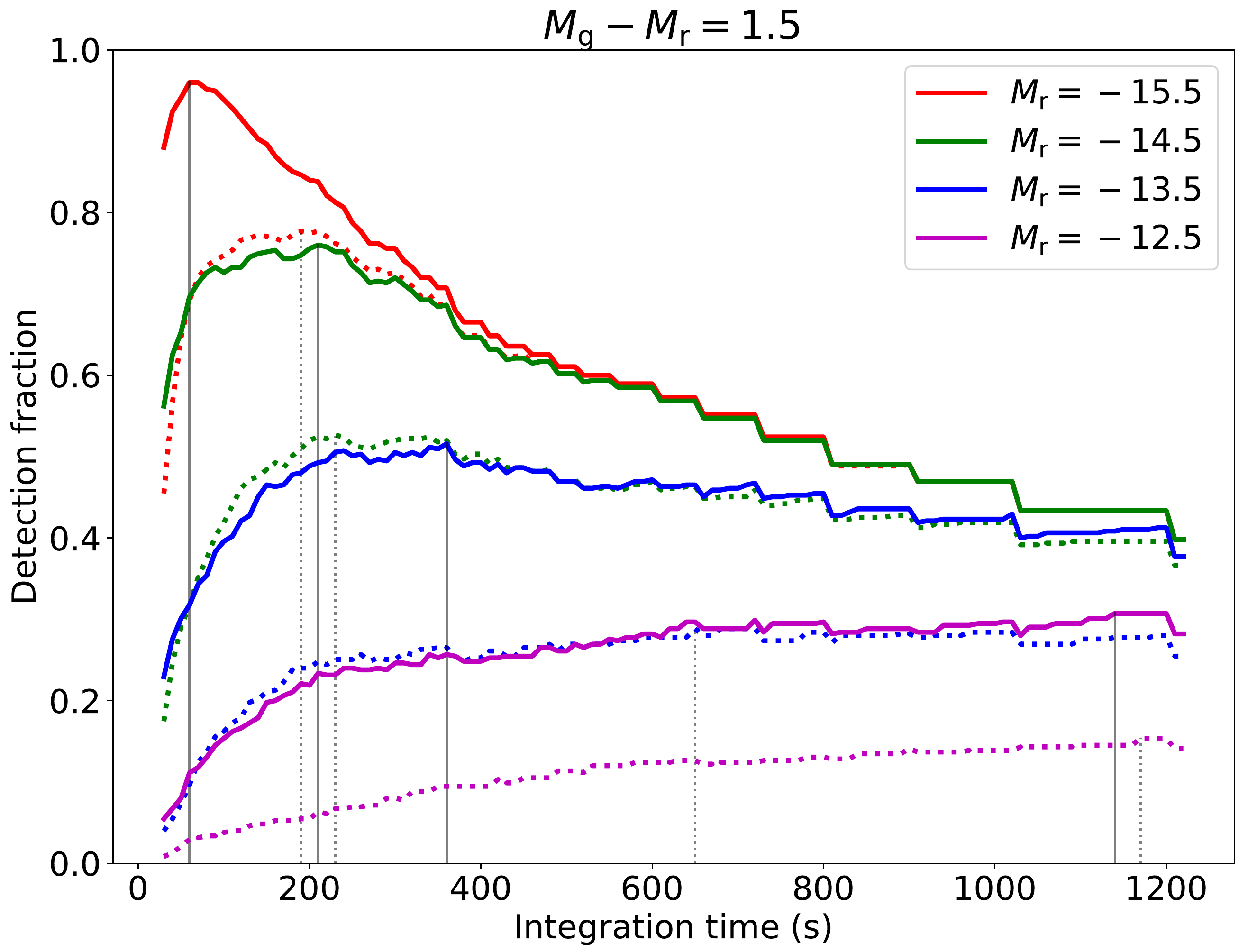}
\caption{The detection fraction as a function of the integration time per pointing. The total observation time is two hours. Based on results obtained from \cite{Rosswog:2016dhy} we chose $M_g - M_r = 1.5$. We present the results for four model light curves, $M_r = -12.5$ (in magenta), $-13.5$ (in blue), $-14.5$ (in green) and $-15.5$ (in red). The dotted lines are for corresponding $M_g = M_r + 1.5$. The vertical black lines denotes the integration time required to reach maximum detectability. If the maximum is not reached during the observation then we set the integration time maxima at 20 minutes. If kilonovae are intrinsically dim, observing deeper improves detection efficiency, while for models that allows for brighter light-curves, we observe that  a detectability-maxima is reached in 100-300 seconds of observation.}
\label{fig:detectionFraction_2hours}
\end{figure}

\begin{figure}[ht]\centering
\includegraphics[width=\linewidth]{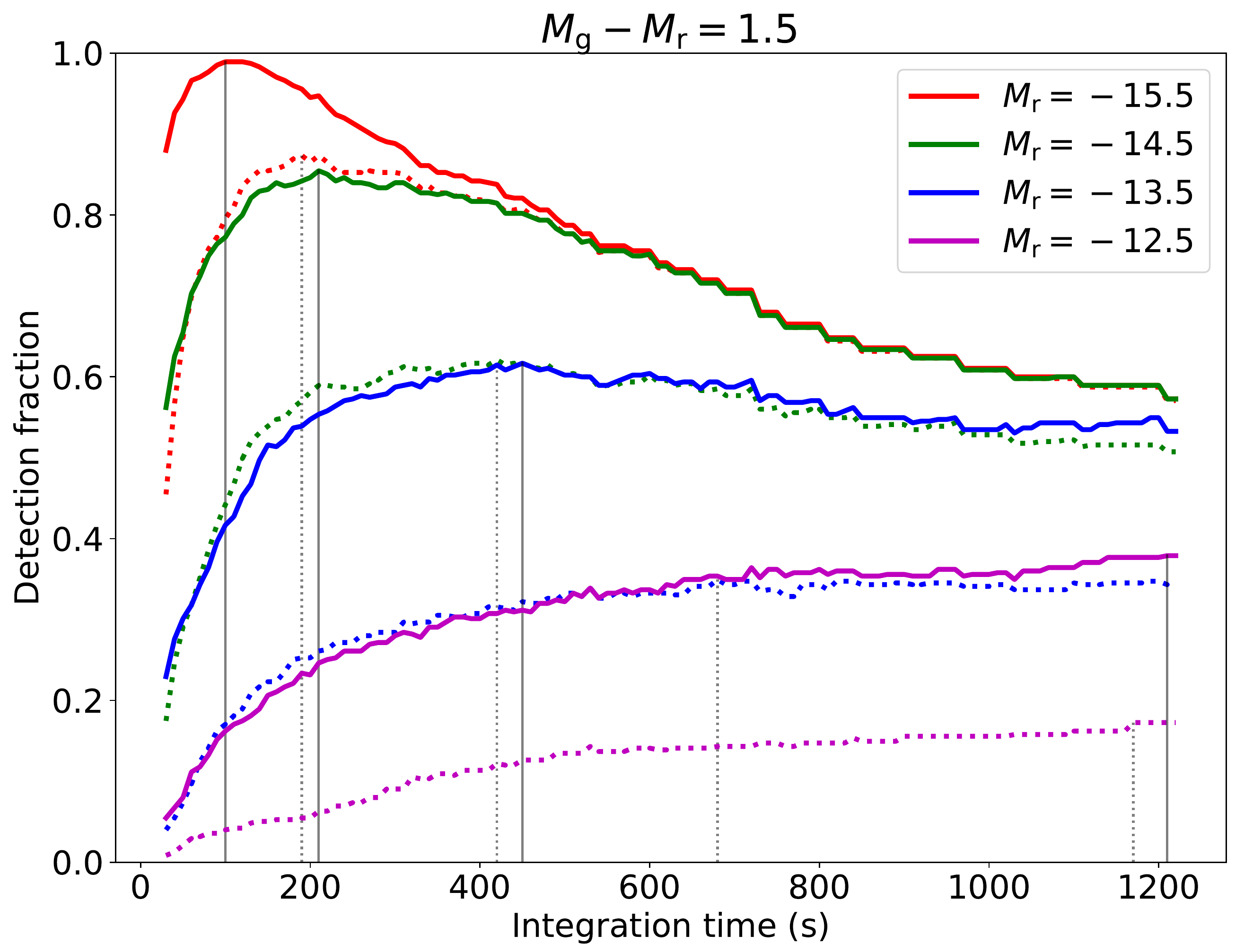}
\caption{The detection fraction as a function of the integration time per pointing. The total observation time is four hours. The color and the style of the lines follows the convention set in Figure \ref{fig:detectionFraction_2hours}.}
\label{fig:detectionFraction_4hours}
\end{figure}

\begin{figure}[ht]\centering
\includegraphics[width=\linewidth]{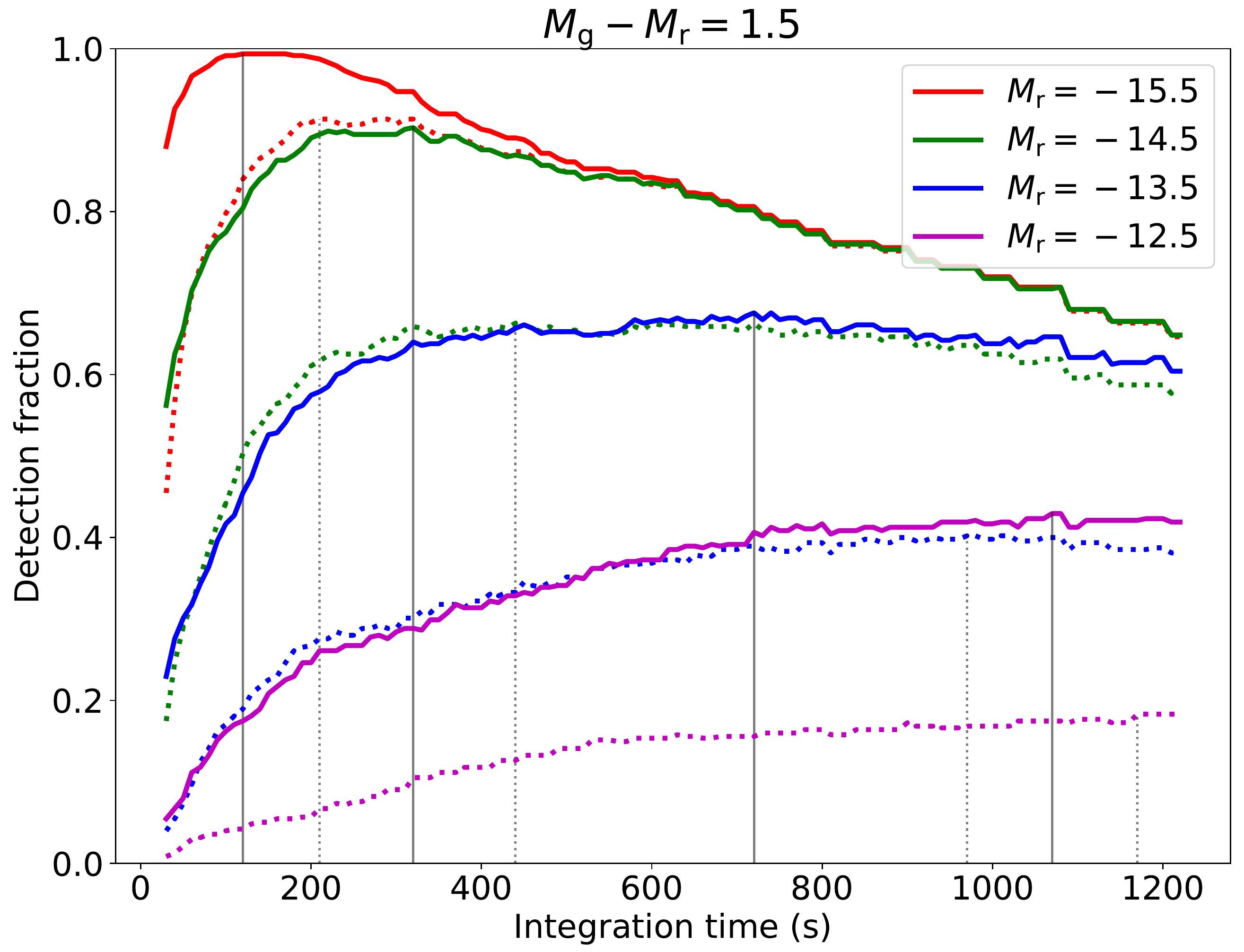}
\caption{The detection fraction as a function of the integration time per pointing. The total observation time is six hours. The color and the style of the lines follows the convention set in Figure \ref{fig:detectionFraction_2hours}.}
\label{fig:detectionFraction_6hours}
\end{figure}

We have restricted our analysis until now to coverage of sky-localization regions. Given the large size of the GW sky-localizations, covering the $90\%$ confidence interval in general could be a challenging task on its own. However, any statement on detectability of EM-counterpart also needs to incorporate discussion of depth of the observation. For the present analysis we use the PTF integration time - limiting magnitude data as a proxy for the ZTF analysis; while ZTF may go somewhat deeper in the same observing time due to an improved detector, assuming the PTF characteristics will be a conservative assumption. We show in Fig. \ref{fig:time_limiting_mag} the relationship between these two quantities 
Note that PTF uses g-band and R-band magnitudes. We converted the R-band magnitude to r-band magnitude by using conversion relation provided by Lupton (2005)\footnote{https://www.sdss3.org/dr10/algorithms/sdssUBVRITransform.php}. If we know the distance of the injected BNS sources and their (intrinsic) absolute magnitudes, we can compute their expected apparent magnitude. Sources whose apparent magnitudes are lower than the limiting magnitude are detectable if the source location is covered by the tiles. In this work, we used a set of absolute magnitude models for kilonovae from \cite{Rosswog:2016dhy}. Specifically, we noted that an absolute magnitude difference of $\sim 1-2$ between $M_g$ and $M_r$ is predicted theoretically for early-time kilonovae light-curves. Thus, in our analysis, we did the following:
\begin{itemize}
    \item We considered $M_g - M_r = 1.5$ and chose four values of $M_r$ : ($-15.5, -14.5, -13.5, -12.5$)
    \item Assumed three total observation times of two hours, four hours, and six hours, motivated by typical observation duration that may be available to observers in a given night of observation. These correspond to Fig.(\ref{fig:detectionFraction_2hours}), Fig.(\ref{fig:detectionFraction_4hours}) and Fig.(\ref{fig:detectionFraction_6hours}) respectively.
    \item Scheduled the observation based on the rank, starting with the highest ranked tile until total observing time is exhausted or the location of the event is found (i.e., the event location is within the ranked tiles that have been covered).
    \item The time spent per tile was progressively increased from 30 seconds (the shortest expected ZTF exposure) to 20 minutes.
\end{itemize}
The source is considered detected when the source location is covered by and observed at a depth greater than what is required to reach the source at the injected distance (for the given model of absolute magnitude). Since, too large of an integration time would not allow us to cover the localization area efficiently, and too wide an attempted coverage would prevent us from reaching the required observation depth, there may exist an optimal integration time for such observation. The goal of this study is to search for this optimal integration time for observation of GW triggered kilonovae with expected brightness. 
 In Figs. \ref{fig:detectionFraction_2hours}, \ref{fig:detectionFraction_4hours} and \ref{fig:detectionFraction_6hours} we show the results for all the cases 

\begin{figure}[ht]\centering
\includegraphics[width=\linewidth]{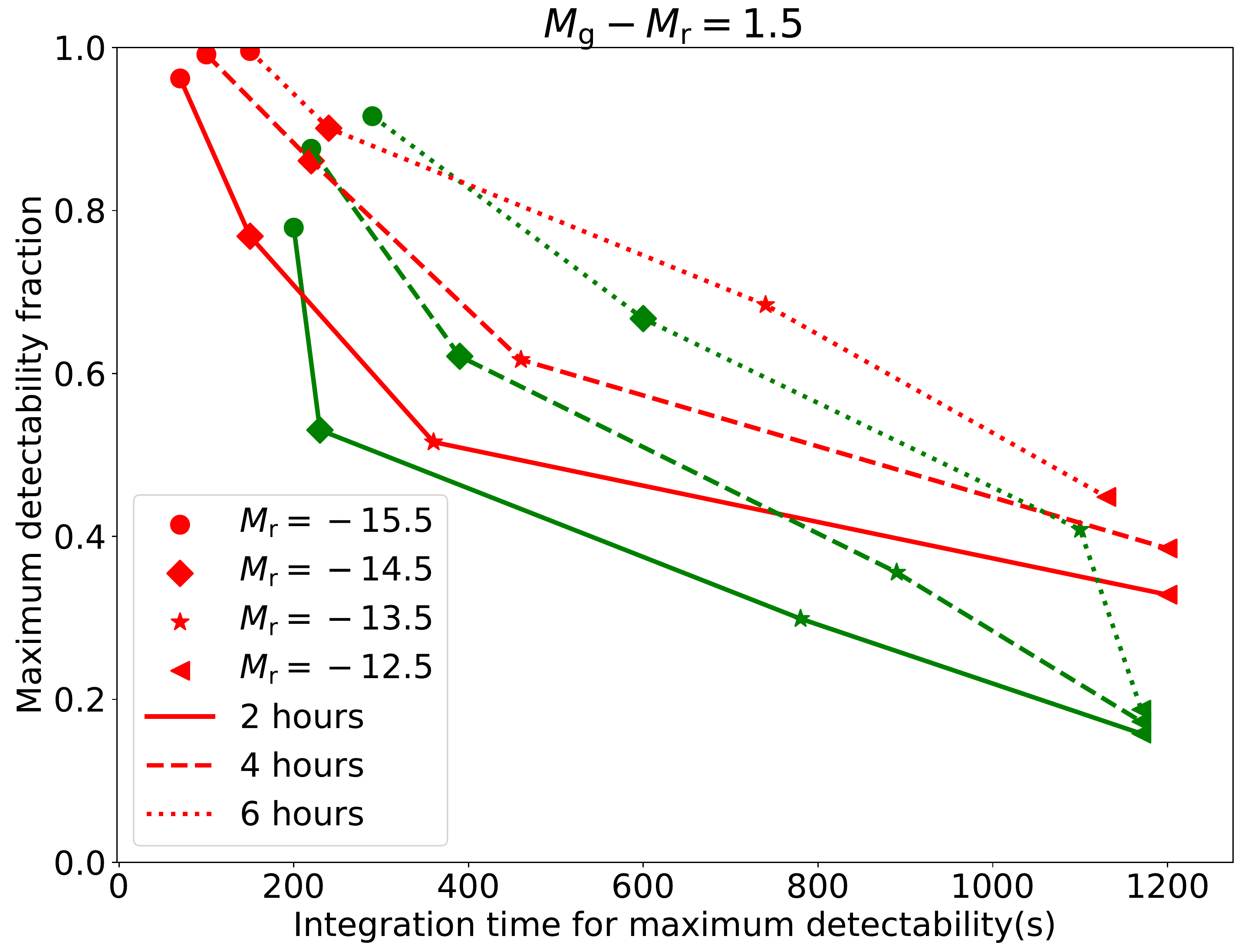}
\caption{The maximum detectability fraction as a function of the integration time required to reach this detectability fraction. The red and green lines are for $r-$band filters and $g-$band filters respectively. The solid, dashed and dotted lines show the variation of the maximum detectability as a function of integration time for two, four and six hours of observations respectively. The circle marker shows the maximum detectability for $M_r = -15.5$, diamond marker shows the same for $M_r = -14.5$ and the star and triangle markers shows the maximum detectability for $M_r = -13.5$ and $-12.5$ respectively, with the corresponding values for $M_g$ obtained by setting $M_g = M_r + 1.5$.}
\label{fig:detectionFraction_Maximas}
\end{figure}

We observe that if kilonovae are intrinsically brighter, we will be able to observe the sky-localizations with lower integration time ($\sim 100-300$ seconds) to achieve maximum detection efficiency. With increasing total observation time, we can afford to increase the integration time to get to higher detection efficiencies that were inaccessible for observation scheduled for shorter total times. This can be seen in the right-ward progression of the peak of each curve from Fig \ref{fig:detectionFraction_2hours} to Fig. \ref{fig:detectionFraction_6hours}. If kilonovae are intrinsically weaker, then we need to invest longer time per pointing, for any given total observation time, to reach a detectability-maxima. For the dimmest models in our experiment, we did not find a maxima within 20 minutes of per-pointing integration time. For such cases, we constrained the maxima to the detection efficiency value corresponding to integration time of 20 minutes. In Fig. \ref{fig:detectionFraction_Maximas} we summarize the results of this analysis, where we plot the maximum detectability fraction as a function of the integration time required to reach maximum detectability for all the cases discussed above.
Thus, we note that a minimum of $\sim 100$ seconds of integration time is essential for ZTF to achieve maximum detectability of the counterparts of GW if kilonovae are intrinsically bright. Greater integration times are required to achieve the detection efficiency maxima if kilonova brightness is more modest. 

In this study we have made two very simplifying assumptions (other than the assumptions on the absolute brightness of the sources). Firstly, we did not include any light-curve evolution. This can be easily incorporated. However, given the time-scales ($\leq 6$ hours) we are considering in your analysis, including more information about the light-curves which varies in many hours to days time-scale does not seem very essential. Secondly, once again we assumed no constraints based on day-night or part of the sky visible from a given location. We also did not consider that the sky-localization itself will be moving across the sky over the course of the observation. These will be addressed in our future studies.

\subsection{Adaptive exposure}
\label{Sec:adaptiveExp}
In the Sec. \ref{Sec:DepthCoverage} we discussed the detectability of the sources for various (static) light-curve models and duration of observations. We varied the integration-time from 60 seconds to 1200 seconds. The general observation was that increasing the integration improves the detectability of kilonovae up to a maximum value and then further increase in integration time results in lower detection fraction. This is due to the fact that for a given duration of observation, the tile coverage of the localization region is inversely proportional to the integration time. If the gravitational wave sky-localization regions are consistent with the true location of the source, then one can adapt the exposure time per tile based on the localization probability. Such investigations had been conducted in the past, for example in \cite{Chan:2015bma}, where the goal was to optimize the number of observing tiles and their  integration time by maximizing detection probabilities. However, given the complexity of the problem such rigorous optimization requires significantly large computation time ($\sim$ days). Here we are looking for a reasonable solution that can be computed in real-time. It was shown in  \cite{Coughlin:2016vif} that integration time proportional to $2/3$rd power of the enclosed probability improves detection probability. In \cite{Salafia:2017ebv} the authors did a more rigorous study that also takes into account the {\em a priori} information about the EM counterpart to decide upon the time to be spent on each pointing.
We also conducted studies on scaling of integration time based on the probability of the ranked-tiles to investigate if the detection efficiency improves. We added a  constraint that the lowest integration time is not lower than 60 seconds (more than the minimum exposure plus slew-time for ZTF) and not higher than 1200 seconds (below sky-dominated regime). The integration time for the $i-$th tile is computed as 
\begin{equation}
\label{Eq:intTimeDist}
t^{i}_{\rm int} = \frac{f(p^{i}_{\rm tile})}{\sum_{i}f(p^{i}_{\rm tile})}\times T_{\rm duration}\,,
\end{equation}
where, $p^{i}_{\rm tile}$ is the probability of the $i-$the ranked-tile, $T_{\rm duration}$ is the total amount of time available for the observation and the factor before $T_{\rm duration}$ is the weight for $i-$th ranked-tile. The function $f$ has a domain $0<f(x)<\infty$ for range $0<x<\infty$ and is an increasing function of $x$.

\begin{figure}[ht]\centering
\includegraphics[width=\linewidth]{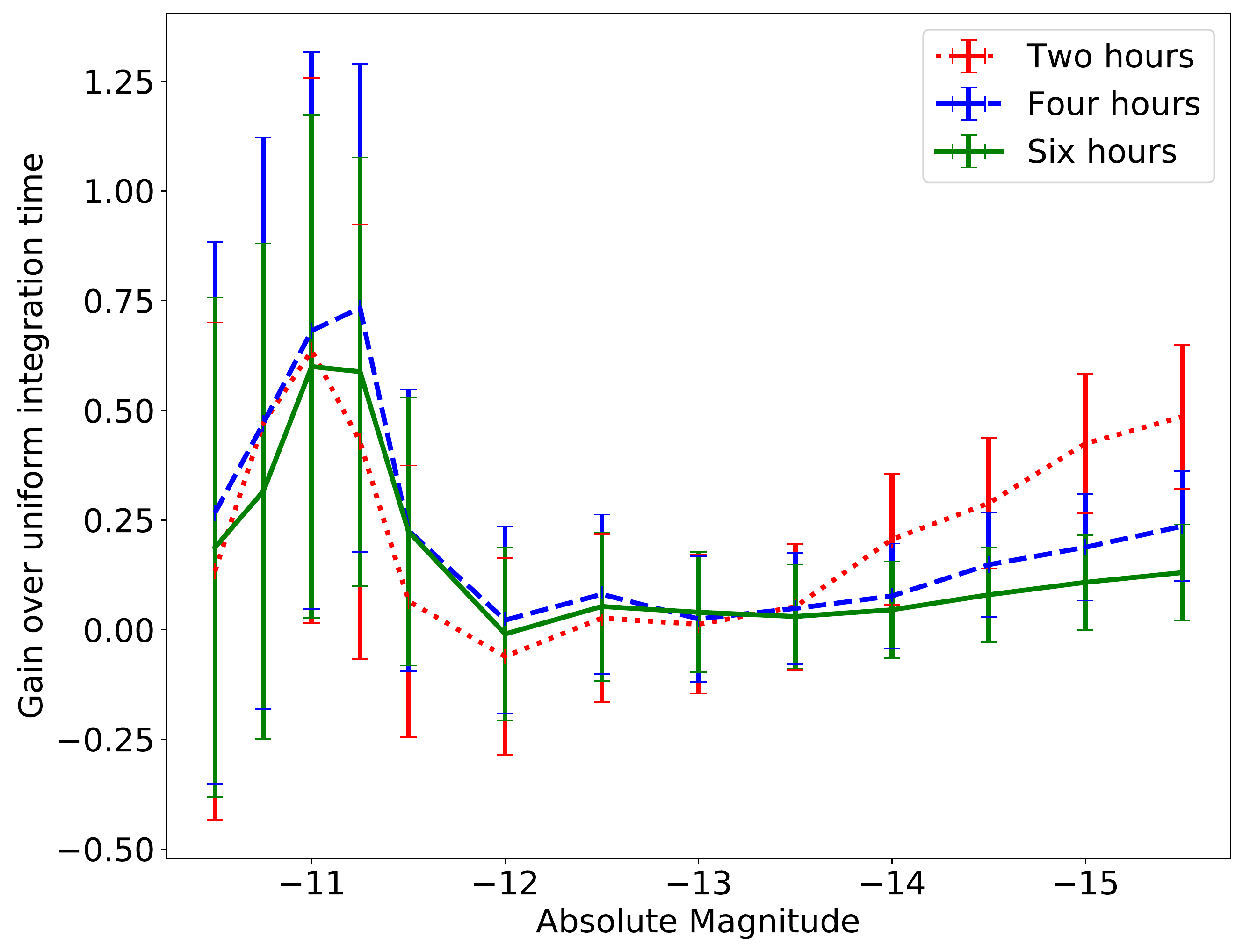}
\caption{Gain in detection for linear adaptation. The gain is defined as the fractional increase (over uniform integration time) in the number of events that were detected using linear adaptation. The solid green, dashed blue and dotted red line shows the gain over uniform integration time for six, four and two hours of observations respectively.}
\label{fig:detectionFraction_LinAdapt}
\end{figure}

The simplest adaptation one can construct is a linear one, where the time spent in the tile is directly proportional to the probability of the tile, $f(p^{i}_{\rm tile}) = p^{i}_{\rm tile}$. We also investigated the detectability for cases where the adaptation uses square-root and the cube-root of the probability value to construct the weights.  The functional form of the weight seems to have some effect with no generic trend. More studies on the exact form of the weight function are being carried out. However, here we present the results of linear-adapt. We conducted the study for a number of absolute magnitudes $(M = -10.5, -11.0, -11.5, \dots, -15.5)$. Three total duration were chosen, two, four and six hours for each case.
Our studies reveal that given the sky-localization areas we expect from binary coalescences, adapting the integration time based on the ranked-tile probability is going to be almost always at least as good as a uniform integration-time of 600 seconds. In Fig. \ref{fig:detectionFraction_LinAdapt} we show the evolution of the gain in detection fraction upon switching to linear adaptation of integration time from a uniform integration time as a function of the absolute magnitude of the sources. We note that for dim and bright kilonova light curves the gain in the detection fraction is the highest. For the brightest models a smaller integration time per tile can lead to detection of the sources. The uniform 600 seconds of integration time is much greater than what is required for most of the cases. Adapting with linear probability weight allows us to reach to a greater number of tiles than it is possible with the uniform 600 seconds of integration time. Thus a greater detection fraction is achieved with the linear adaptation. For very weak kilonova light curves, most systems are undetectable with 600 seconds of integration time. Such systems become detectable with the linear adaptation if a larger integration time is allotted to the tile containing the source. If gravitational wave sky-localization regions are consistent with the actual location of the sources, then using the localization probability as a measure of the amount of time spent on the tile is almost always profitable.

In Table \ref{tab:intTimeOptim}, we compare the number of sources detected (covered and depth reached), for the various light-curve models, for uniform integration time and linearly adapted integration time. Note that these results were obtained for a specific instance of linear-adapt, namely $f(p_{\rm tile}^i) = p_{\rm tile}^i$. Thus, the resulting distribution of integration times per tiles is not unique, and in principle there are an infinite number of ways to construct the per tile integration times. There could exist a particular functional form for adapting the integration time that may depend upon the telescope FOV and integration time dependence of limiting magnitude. In our limited studies, we found that for integration times $\propto p_{\rm tile}^{\alpha}$ where $0.0 < \alpha \leq 1.0$, we get better detection efficiency than uniform integration time of 600 seconds. Stiffer dependence of integration time on the probability of the tiles turned out to be unproductive in our studies.

\begin{table}[hbt]
\caption{Comparison of the number of events detected (out of 475) using uniform integration time and a linearly adapted integration time}
\label{tab:intTimeOptim}
\centering
\begin{tabular}{llll}
\toprule
\toprule
\cmidrule(r){1-2}
$M_r$ & Detected (uniform) & Detected (linear-adapt)\\
\midrule
$-10.5$ & $11$ & $15$\\
$-11.5$ & $38$ & $46$\\
$-12.5$ & $92$ & $106$\\
$-13.5$ & $187$ & $191$\\
$-14.5$ & $248$ & $288$\\
$-15.5$ & $253$ & $337$\\
\bottomrule
\end{tabular}
\label{tab:uniform_vs_adaptive}
\end{table}

\section{Discussion}
\label{Sec:discussion}
In this work, we presented the tiling strategy that would be best suited for ZTF for follow-up of GW sky-localizations. The solution given by ranked tiling holds even if the fixed-tiles constraint is relaxed (as we saw with the case of two set cases).  We also implemented the ranked-tiling algorithm for detection-strategy. The detection strategy has two components, the tiling strategy, and the exposure strategy. The exposure strategy involves a judicious choice of the amount of time we should spend in each ZTF field in order to maximize the chance of detection. We found that if we linearly scale the exposure time based on the gravitational-wave probability of the tile we can improve our counterpart detection efficiency. 
These results could be further improved by implementing distance prior in the optimization. Moreover, the gravitational-wave sky-localization from the second observing run would also contain the distance information \cite{2041-8205-829-1-L15} of the source which would provide us a more natural way of defining the weighted volume. We will perform studies on such sky-maps in the future.

We make the code-base for ranked tiling and adaptive-integration time available online \footnote{\url{https://github.com/shaonghosh/sky_tiling}}. The repository also contains example pixel-tile maps for some chosen telescopes, that are required for rapid computation of the ranked tiles. The general user can create their own pixel-tile maps suited for their telescope field-of-view using the scripts present in the repository.

We are exploring the possibility of the inclusion of an $i$-band filter for ZTF. It is expected that kilonova light curves are brighter in the infra-red and hence the detection efficiency could significantly improve.

Finally, we plan to also include galaxy-catalog information in our observation strategy. Currently, the ranking of the tiles are determined solely from GW sky-localization. However, we will compute the mass of the galaxies within each tile to weigh the probability enclosed within its boundary. We can further modify our ranking statistic by including spectral and/or photometric information from the galaxies within the catalog. We can include Milky-way dust maps to incorporate foreground absorption in our ranking.

\phantomsection
\section*{Acknowledgments} 
The authors will like to thank Eric Bellm and David Cook for providing the ZTF/PTF specification data which was used for this analysis. Most of the analysis that is presented in this paper was carried out on the Coma Computing Cluster of the Radboud University, Nijmegen, the Netherlands. SG and PRB acknowledge NSF Award PHY-1607585 that funded this work. DLK acknowledges the GROWTH project funded by the National Science Foundation under Grant No 1545949 which supported this work.

\addcontentsline{toc}{section}{Acknowledgments} 

\phantomsection
\bibliographystyle{unsrt}
\bibliography{references}


\end{document}